# Large Cross-phase Modulation Based on Double EIT in a Four-level Tripod Atomic System

Shujing Li, Xudong Yang, Xuemin Cao, Chunhong Zhang, Changde Xie, Hai Wang[*]

*The State Key Laboratory of Quantum Optics and Quantum Optics Devices, Institute of Opto-Electronics, Shanxi University, Taiyuan, 030006, People's Republic of China*

We report the experimental observations on the simultaneous EIT effects for probe and trigger fields (double EIT) as well as the large cross-phase modulation (XPM) between the two fields in a four-level tripod EIT system of the D1 line of $^{87}$Rb atoms. The XPM coefficients (larger than $2\times10^{-5}$ cm$^2$/W) and the accompanying transmissions (higher than 60%) are measured at slightly detuning of the probe field from the exact EIT resonance condition. The presented system can be applied in the recently proposed quantum information processing with weak cross-Kerr nonlinearities.

*PACS:* 42.50.Gy, 32.80.–t, 42.65.-k

Large optical cross-phase modulation (XPM) or cross-Kerr nonlinearity is essential to demonstrate quantum phase gates [1-3] and generate quantum entanglement of single photons [4]. Recent studies have shown that the weak cross-Kerr nonlinearity can also be used for constructing a near deterministic controlled-Not gate [5], performing a nondestructive Bell-state detection [6], and implementing a new approach for quantum computation[7]. However, the absence of sufficient XPM in conventional media becomes an obstacle of its applications in quantum information processing(QIP). Electromagnetically induced transparency (EIT) technology would be a promising avenue for solving the problem [2-4]. The large cross-Kerr phase shift in four-level N-type EIT system has been theoretically proposed [8] and experimentally demonstrated in cold atoms[9]. However, in the N-type EIT system, the large cross-phase modulation between probe and trigger (signal) pulses may not occur due to a group velocity mismatch between the two pulses[10-12]. Recently, several schemes have been proposed to obtain large XPM phase shift in various multi-level systems[2-3,11-12]. All of the proposed schemes are based on the simultaneous EIT for two weak fields (named double EIT), that is because in this case both group velocities of the two weak fields are small and comparable, so long interacting time for producing large XPM can be obtained. Some related studies, such as the dark resonance switching [13], nonlinear optics in four- or five-level EIT systems[14-15] have been experimentally demonstration and a Kerr effect enhancement scheme based Raman gain[16] has been proposed recently. Up to now, the experimental demonstration of the large XPM effect between two weak fields based on double EIT has not been presented to the best of our knowledge. Recently, a variety of theoretically proposals on the applications of weak cross-Kerr nonlinearities [5-7] have been proposed. The experimental search for the effect is being expected anxiously. Under the motivation we accomplished the presented experimental study.

In this letter, we report the first experimental observations on the simultaneous EIT for probe and trigger (double EIT) fields as well as the enhanced cross-phase modulation between the two fields in a four-level tripod EIT system of the D1 line of $^{87}$Rb atoms. The relevant atomic levels are shown in Fig.1(a). The probe field $E_P$ of frequency $\omega_p$ is left-circularly-polarized ($\sigma^-$), with a Rabi frequency $\Omega_p = \mu E_p/\hbar$, coupling to the transitions from levels |a$_{i+1}$> to |e$_i$> (i=1,2). The coupling field $E_C$ of frequency $\omega_C$ is also $\sigma^-$-polarized with a Rabi frequency $\Omega_C = \mu E_c/\hbar$ to drive the levels |b$_{i+2}$> to |e$_i$> transition (i=1,2,3). The trigger beam $E_T$ of frequency $\omega_T$ is right-circularly-polarized ($\sigma^+$), with a Rabi frequency $\Omega_T = \mu E_T/\hbar$, coupling to the levels |b$_i$> to |e$_i$> transitions (i=1,2,3). $\mu$ is the dipole moment for $^{87}$Rb D1 transitions. In this case, the system is coherently prepared into two four-level tripod-type systems [17], one is formed by the levels |a$_2$>-|b$_3$>-|b$_1$>-|e$_1$> (S1) and another by the levels |a$_3$>-|b$_4$>-|b$_2$>-|e$_2$> (S2). The $\sigma^-$ probe transition and $\sigma^+$ trigger transition share a common excited state |e$_1$> (|e$_2$>), which induce a coupling between the probe and trigger fields in such a coherently prepared tripod system, so the cross-Kerr phase modulation between the two fields will be enhanced. The total probe (trigger) susceptibility $\chi_p$ ($\chi_T$) should include the contributions of both systems S1 and S2, that is $\chi_p = \chi_{P1} + \chi_{P2}$ ($\chi_T = \chi_{T1} + \chi_{T2}$). In the presented measurement of the XPM between the probe and trigger beams, the condition of $|\Omega_{P,T}|^2 \ll |\Omega_C|^2$ is satisfied. Solving the density-matrix equations of the Eq.3 in Ref.[17] under the steady-state condition[2-3], the susceptibilities $\chi_{pi}$ (i=1,2) and $\chi_{Ti}$ (i=1,2) are obtained:

$$\chi_{Pi} = -\frac{N|\mu_{ai+1,ei}|^2}{\hbar\varepsilon_0}\left\{\frac{\rho_{ai+1,ai+1}-\rho_{ei,ei}}{-\tilde{\Delta}_P+|\Omega_{Ti}|^2/4\tilde{\Delta}_{PT}+|\Omega_{Ci}|^2/4\tilde{\Delta}_{PC}} - \frac{|\Omega_{Ti}|^2(\rho_{bi,bi}-\rho_{ei,ei})/4}{\tilde{\Delta}_{PT}(-\tilde{\Delta}_T^*+|\Omega_{ci}|^2/4\tilde{\Delta}_{TC}^*)(-\tilde{\Delta}_P+|\Omega_{Ci}|^2/4\tilde{\Delta}_{PC})}\right\} \quad (1a)$$

$$\chi_{Ti} = -\frac{N|\mu_{bi,ei}|^2}{\hbar\varepsilon_0}\left\{\frac{\rho_{bi,bi}-\rho_{ei,ei}}{-\tilde{\Delta}_T-|\Omega_{Pi}|^2/4\tilde{\Delta}_{PT}^*+|\Omega_{Ci}|^2/4\tilde{\Delta}_{TC}} - \frac{|\Omega_{Pi}|^2(\rho_{ai+1,ai+1}-\rho_{ei,ei})/4}{\tilde{\Delta}_{PT}^*(-\tilde{\Delta}_P^*+|\Omega_{ci}|^2/4\tilde{\Delta}_{PC}^*)(-\tilde{\Delta}_T+|\Omega_{Ci}|^2/4\tilde{\Delta}_{TC})}\right\} \quad (1b)$$

where $\tilde{\Delta}_P = \Delta_P + i\gamma_0$, $\tilde{\Delta}_T = \Delta_T + i\gamma_0$, $\tilde{\Delta}_{PT} = \Delta_P - \Delta_T + i\gamma_1$, $\tilde{\Delta}_{PC} = \Delta_P - \Delta_C + i\gamma_2$, $\tilde{\Delta}_{TC} = \Delta_T - \Delta_c + i\gamma_3$, $\Delta_P$, $\Delta_C$ and $\Delta_T$ are the probe, coupling and trigger frequency detuning[17], respectively. $\gamma_j$ (j=0,1,2,3) describes decay of populations and coherences. $\rho_{ai,ai}$, $\rho_{bi,bi}$ and $\rho_{ei,ei}$ respectively are the populations of states $|a_i\rangle$, $|b_i\rangle$ and $|e_i\rangle$. $\Omega_{Pi} = \Omega_p C_{ei,ai+1}$, $\Omega_{Ci} = \Omega_C C_{ei,bi+2}$ and $\Omega_{Ti} = \Omega_T C_{ei,bi}$ are the probe, coupling and trigger Rabi frequencies for the transitions from Zeeman levels $|a_{i+1}\rangle$ to $|e_i\rangle$, $|b_{i+2}\rangle$ to $|e_i\rangle$ and $|b_i\rangle$ to $|e_i\rangle$, respectively. $C_{ei,bj(ak)}$ is a coefficient related to the dipole moment $\mu_{ei,bj(ak)}$ for the transition from levels $|b_j\rangle$ ($|a_k\rangle$) to $|e_i\rangle$ with a expression $\mu_{ei,bj(ak)} = C_{ei,bj(ak)}\mu$.

Figure 1(b) depicts the experimental setup. LD1 (for probe beam) and LD2 (for coupling and trigger beams) are the frequency-stabilized diode lasers (linewidths~1MHz) with the grating feedback. The LD2 laser beam is split into two parts by a beam splitter (BS1), one of them serves as the trigger beam and the other one serves as the coupling beam. The trigger beam passes through an AOM system for scanning its frequency around $\omega_{be}$ (see the Ref. [17] for the details). The EIT dispersion curves of the probe and trigger beams are measured with a common Mach-Zehnder (M-Z) interferometer. The M-Z interferometer consists of two beam displacing polarizers, BD1 and BD2. The linearly-polarized input probe beam (with a chosen polarization angle) is separated into two orthogonally polarized beams, s- and p-polarized beams, by BD1. They are used for the probe and the probe reference beams, respectively. The two beams propagate parallel along line 2 ($L_2$) and line 1 ($L_1$), respectively. Similarly, the linearly-polarized input trigger beam is also separated into orthogonal s- and p-polarized output beams by BD1, which serve as the trigger reference and the trigger beams, respectively. The p-polarized trigger beam propagates along line 2 and overlaps with the s-polarized probe beam. The s-polarized trigger reference beam propagates along line 3 ($L_3$), which is parallel with line 2. Both of the spacial distances between line 1 and 2, as well as between line 2 and 3 are 4mm. The s-polarized coupling beam propagates through a Rb cell with a small angle (~ 1°) relative to the line 2 and overlaps with the probe and trigger beams inside the Rb cell. The s-polarized probe and coupling beams become $\sigma^-$-polarized while the p-polarized trigger beam becomes $\sigma^+$-polarized after they respectively pass through a λ/4 wave-plate. Since the three laser beams co-propagate inside the Rb cell, the first-Doppler broadening in two-photon frequency detunings ($\Delta_P - \Delta_T, \Delta_P - \Delta_C$ and $\Delta_C - \Delta_T$) will be eliminated [17]. The length of the atomic cell is $l$=50 mm with magnetic shielding and its temperature is stabilized to about 63.5℃. A weak magnetic field (~150mG) in the z direction of the atomic cell (parallel to the line 2) is applied by means of a Helmholtz coils to provide a quantization axis. The 1/e intensity diameters of the probe, trigger and coupling beams are about 1mm, 1mm and 3mm at the center of the cell, respectively. The probe and trigger beams are regained to their original linear polarizations after respectively passing through another λ/4 wave-plate. The intensities of the probe, trigger, probe reference and trigger reference beams are detected by detectors D3, D2, D1 and D4, respectively. The transmitted probe (s-polarized) and probe reference (p-polarized) beams from BS2, overlap each other at the exit of BD2 to be combined into a beam P again. Similarly, the trigger (p-polarized) and trigger reference (s-polarized) beams are combined into another beam T at the exit of BD2. Then, P (T) beam passes through a MgO:LiNbO$_3$ crystal Li1 (Li2) and a λ/2 wave-plate. Both p- and s-polarized components of P (T) beam are rotated 45° by the λ/2 wave-plate. This combined P (T) beam is split into two parts P1 (T1) and P2 (T2) with an equal power by PBS1 (PBS2). The detectors D7 and D8 as well as D5 and D6 form two homodyne detectors H1 and H2, respectively, whose local phase can be set to $\pi/2$ by changing the voltage applied on Li1 and Li2 at EIT resonance. In this case, the differential signals [18] $\Delta I_{H1} \propto 2|E_{RP}||E_P|e^{-\alpha_P(\omega)l/2}kn_P(\omega)l$ and $\Delta I_{H2} \propto 2|E_{RT}||E_T|e^{-\alpha_T(\omega)l/2}kn_T(\omega)l$ from H1 and H2 will give the probe dispersion $kn_P(\omega)l$ and trigger dispersion $kn_T(\omega)l$, where $E_{RP}$ ($>>E_P$) and $E_{RT}$ ($>>E_T$) are the probe and trigger reference fields, respectively.

We observed the EIT windows for probe and trigger fields when coupling detuning $\Delta_C = 0$. The powers of the coupling, probe and trigger beams are set to $P_C=14mW$ ($\Omega_C \approx 70MHz$), $P_P=8\mu W$ ($\Omega_P \approx 3MHz$) and $P_T=10\mu W$ ($\Omega_T \approx 3MHz$), respectively. When the coupling beam was on, we scanned the probe frequency across $\omega_{ae}$ to measure the probe absorption spectrum at $\Delta_T = 0$. A EIT window (Fig.2(a)) for the probe beam with a linewidth of ~2MHz appears at the resonance of $\Delta_C = \Delta_P = 0$, which derives from the three-level $\Lambda$ system |b3>-|e1>-|a2> (|b4>-|e2>-|a3>). At $\Delta_P = 0$ scanning the trigger frequency around $\omega_{be}$ with the AOM system [17], the trigger EIT signal (Fig.2(b)) at the resonance of $\Delta_C = \Delta_T = 0$ with a linewidth of ~2MHz was observed, which derives from another three-level $\Lambda$ system |b3>-|e1>-|b1> (|b4>-|e2>-|b2>).

We also observed the cross-phase modulation between the probe and trigger fields under the conditions of different powers for the two beams. The measured results are shown in Fig.3. During the measurements of Fig.3, the coupling beam with a power of $P_C=14mW$ was always on and the probe frequency was scanned across $\omega_{ae}$. First, under the condition of $P_p=8\mu W$ ($\Omega_P \approx 3MHz$) and $P_T=300\mu W$ ($\Omega_T \approx 18MHz$), i.e., $\Omega_P < \Omega_T (<< \Omega_c)$, we measured the modulation of the probe field by the trigger field. The curves I of Fig.3(a) and I' of Fig.3(b) respectively present the probe EIT absorption $\alpha_P(\omega)l$ and dispersion $kn_P(\omega)l$ as the function of $\Delta_P$ when the trigger beam is blocked, which derive from the three-level $\Lambda$-type systems. When the trigger beam is applied, the four-level tripod systems (S1 and S2) are formed and the EIT absorption (curve II of Fig.3(a)) and dispersion (curve II' of Fig.3(b)) from the tripod systems become much larger than the corresponding results from the $\Lambda$-type systems, respectively. From Fig.3(b), we can see that the two EIT dispersion peaks of the probe beam occur at $\Delta_P \approx -0.7 MHz$ and $\Delta_P \approx 0.6 MHz$, respectively. At the left dispersion peak, a negative XPM phase shift $\Psi_P^N$ of ~-2.5° was achieved, with a transmission of 70%. Next, we measured the modulation of the weaker trigger beam with a power of *10μW* by the stronger probe beam with a power of *300μW*. Traces (i) of the Fig.3(c) and (i') of the Fig.3(d) present the trigger EIT absorption $\alpha_T(\omega)l$ and dispersion $kn_T(\omega)l$ signal when the probe beam is blocked, which correspond to the absorption and dispersion at resonance $\Delta_C = \Delta_T = 0$, respectively. Curves ii of Fig.3(c) and ii' of the Fig.3(d) are the measured trigger EIT absorption and dispersion, respectively, when the probe beam is applied. The top of EIT in curve ii of Fig.3(c) (four-level tripod systems) is far beyond the trace i of Fig.3(c) (three-level systems) when both the probe and trigger fields are tuned to the dark states ($\Delta_C = \Delta_T = \Delta_P$). Simultaneously, a sharp EIT dispersion curve ii' of Fig.3(d) appears at $\Delta_C = \Delta_T = \Delta_P = 0$, which is greatly different with that on the trace i' of Fig.3(d). At $\Delta_P \approx -0.5 MHz$, a maximum of the trigger XPM phase shift $\Phi_T^N$ of ~5° is achieved. The above results show that the maximums of both probe and trigger XPM phase shifts can be achieved at $\Delta_P \approx -0.5 MHz$, but the signs of them are opposite.

Successively, we observed the XPM between the weak probe and trigger fields, both of which have the same power $P_P \approx P_T \approx 14\mu W$ ($\Omega_P \approx \Omega_T \approx 4MHz$). The curves I of Fig.3 (e) and I' of Fig.3 (f) are the measured probe EIT absorption and dispersion without the trigger beam on, the curve i of Fig.3 (g) and i' of Fig.3(h) are the measured trigger absorption and dispersion without the probe beam on. When the two beams were on (Note that the coupling beam with a power of 14mW is always on), we simultaneously measured the EIT absorption curves II (probe) in Fig.3(e) and ii (trigger) in Fig.3(g)), as well as the dispersion curves II' (probe) in Fig.3(f) and ii' (trigger) in Fig.3(h)) in a scanning of $\Delta_p$, respectively. The results show that the modulations of the probe EIT absorption and dispersion by trigger beam are small, but the modulation of the trigger EIT absorption and dispersion by probe beam is quite obvious. At $\Delta_P = \pm 0.5 MHz$, the maximal trigger nonlinear phase shift $\sim \pm 1.4°$ is achieved with an absorption of ~74%. However such a large absorption is not desired in the practical application of QIP.

For exploring the optimal condition to obtain a large XPM and an accompanying small absorption, we measured the XPM phase shifts and the transmissions at the different $\Delta_C$ under the condition of the weak probe and trigger fields ($P_p \approx P_T \approx 14\mu W$). The curves A and B of Fig.4(a) plot the measured XPM coefficients $n_P^{(2)}$ and $n_T^{(2)}$ obtained at $\Delta_P - \Delta_C \approx -0.5MHz$ as the function of $\Delta_C$. The curves C and D of Fig.4(b) plot the simultaneously measured transmissions at $\Delta_P - \Delta_C \approx -0.5MHz$. Increasing $\Delta_C$, the absolute values of XPM coefficients at $\Delta_P - \Delta_C \approx -0.5MHz$ for $n_P^{(2)}$ and $n_T^{(2)}$ go down, but the accompanying transmissions of the probe and trigger beams become larger, which allows people to choice a appropriate coupling frequency detuning to achieve a enough XPM phase shift with a smaller absorption. Fig.4(a) shows that $n_T^{(2)}$ is much larger than $n_P^{(2)}$. This asymmetry of $n_T^{(2)}$ and $n_P^{(2)}$ is not agreement with the prediction in Ref.[2], in which the $n_T^{(2)} \leftrightarrow n_P^{(2)}$ exchange is symmetric. A main reason is that the large differences between the probe Rabi frequencies ($\Omega_{p1} = \Omega_{p2} = 1.15MHz$) and the trigger

Rabi frequencies ($\Omega_{T1} = 2.83MHz$, $\Omega_{T2} = 2MHz$) result in the difference of the populations of two ground state |a$_2$> and |b$_1$> (|a$_3$> and |b$_2$>). The calculated populations with Eq.3 in Ref.[17] are $\rho_{a2,a2} \approx 0.38$, $\rho_{a3,a3} \approx 0.3$, $\rho_{b1,b1} \approx 0.12$, $\rho_{b2,b2} \approx 0.2$ for $\Omega_{p1} = \Omega_{p2} = 1.15MHz$, $\Omega_{T1} = 2.83MHz$, $\Omega_{T2} = 2MHz$. At the same time, we theoretically calculated the fitting curves $n_P^{(2)}$ and $n_T^{(2)}$ as well as the accompanying transmission at $\Delta_P - \Delta_C = -0.5MHz$ with Eq.[1], which are shown in the curves A' and B' as well C' and D'. The theoretical calculations are in reasonable agreement with the experimental results.

In the presented system, the probe (trigger) EIT window always exist when the trigger (probe) field is on or off and at the same time the XPM phase shift can be also acquired. Thus, this system allows us to produce XPM phase shifts with small absorptions. For example, the trigger XPM coefficient can reach $2 \times 10^{-5} cm^2/W$ with a transmission of ~60% (see Fig.4). Such a property is much better than that obtained with the N-type system, in which the probe EIT window will move into an absorption peak when a XPM phase shift is acquired [9]. It is easy to be calculated that if a probe beam with the intensity of ~$0.2mW/cm^2$ ($\Omega_P \approx 1MHz$) is applied, which corresponds to the case that a probe pulse consisting of one photon is tightly focused to a spot size of a half wavelength at 795nm, for a duration of 1$\mu s$, the trigger XPM phase shift induced by the probe field will reach ~0.0016 rad. Such a XPM phase shift may satisfy the requirement [5] of $\sqrt{n}\theta^2 \approx 5$ (the mean photon number n per pulse is on the order of $4 \times 10^{12}$), thus, the presented system can be found practical applications in QIP based on a photon number QND detector[5-7]. Although the conditional phase shift which is defined as $\Phi_P^N + \Phi_T^N$ [1] in QPG still can't reach ~$\pi$ in our experimental conditions, if the laser linewidths of probe and trigger beams are suppressed down to 5kHz, a conditional phase shift ($\Phi_P^N + \Phi_T^N$) based on the XPM between the two pulses with one photon will reach 1.2 rad for $\Omega_C = 30 MHz$, which allows us to perform the QPG operation.

In summary, we have experimentally demonstrated the simultaneous EIT for probe and trigger fields (double EIT) and a large XPM phase shift between the two fields in the four-level tripod EIT systems. The double EIT windows is important for matching the group velocities of the probe and trigger beams for obtaining a large XPM between the two optical pulses [2].

We thank M. Xiao for helpful discussions and acknowledge the funding supports by NSFC (#60325414, 60578059，60736040, 10640420195, and RGC60518001), and 973 Program (#.2006CB921103). *Corresponding author H. Wang's e-mail address is wanghai@sxu.edu.cn.

**Caption:**

FIG. 1. (Color online) (a) Relevant energy diagram of the D1 line in $^{87}$Rb atom. (b) Experimental setup.

FIG. 2. (Color online) The measured simultaneous EIT windows for (a) probe and (b) trigger beams created by a coupling beam with a power of *14mW* at detuning $\Delta_C = 0$.

FIG. 3. (Color online) The measured probe absorption ((a) and (e)) as well as dispersion ((b) and (f)), and trigger absorption ((c) and (g)) as well as dispersion ((d) and (h)) signals as the function of $\Delta_P$ for $\Delta_C = \Delta_T = 0$ and *Pc=14mW*. The black (dotted) curves are only with the coupling beam on. The red (solid) curves in Fig.3 (a), (b), (e) and (f) (Fig.3 (c), (d), (g) and (h)) are with both coupling beam and trigger (probe) beam on.

FIG. 4. (Color online) XPM Coefficients at $\Delta_P - \Delta_C \approx -0.5MHz$ (a) and accompanying transmissions (b) as a function of the coupling frequency detuning, respectively, for $P_P \approx P_T \approx 14\mu W$. Curves (A) and (C) (curves (B) and (D)) are the measured probe (trigger) XPM coefficient and accompanying transmission. Curves (A'), (B') (C') and (D') are corresponding theoretical results with experimental parameters $\gamma_0 = 3.5MHz$ , $\gamma_1 = 0.5MHz$ , $\gamma_2 = 1.5MHz$ , $\gamma_3 = 1.0MHz$ , $\Omega_C = 70MHz$ , $\Omega_P = \Omega_T = 4MHz$ , $N = 3.72 \times 10^{11} / cm^3$.

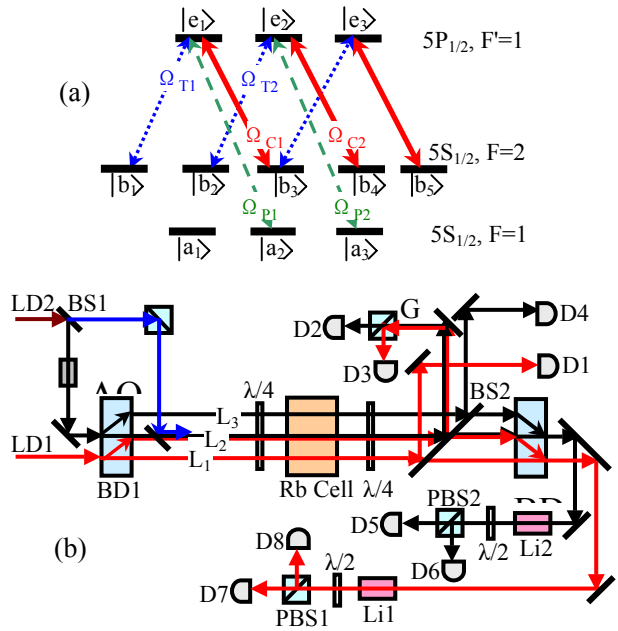

Fig. 1

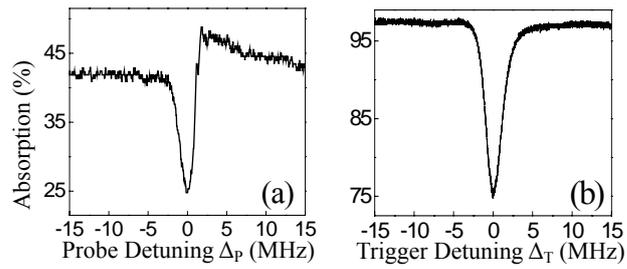

Fig. 2

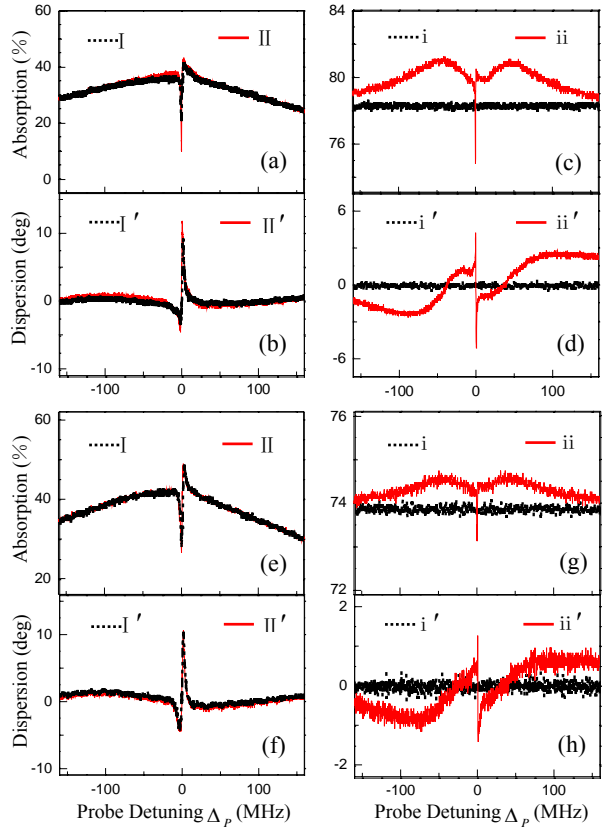

Fig. 3

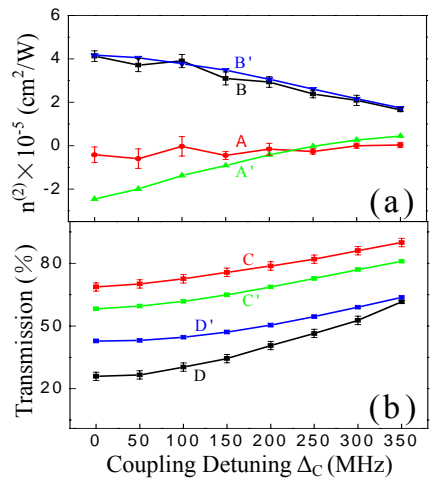

Fig. 4

13